\documentstyle[preprint,prd,aps,here]{revtex}
\tightenlines
\newcommand{\be}{\begin{eqnarray}}
\newcommand{\ee}{\end{eqnarray}}
\newcommand{\bea}{\begin{eqnarray}}
\newcommand{\eea}{\end{eqnarray}}
\newcommand{\ba}{\begin{array}}
\newcommand{\ea}{\end{array}}

\newcommand{\no}{\nonumber}

\newcommand{\p}{\roarrow{p}}
\newcommand{\k}{\roarrow{k}}

\newcommand{\D}{\roarrow{D}}
\newcommand{\gam}{\roarrow{\gamma}}

\newcommand{\Op}{{\cal O}}

\newcommand{\V}{{\cal V}}

\newcommand{\lw}[1]{\smash{\lower1.6ex\hbox{#1}}}
\newcommand{\nabr}{\rlap{\hbox{$\nabla$}}
                   \raise 8 pt \hbox{$\hspace{-0.05cm}\leftarrow$}}
\newcommand{\nabd}{\loarrow{D}}
\newcommand{\vsa}{\rlap{\hbox{$\longrightarrow$}}
                   \raise 7pt \hbox{\scriptsize VSA~~}}
\def\simgt{\rlap{\lower 3.5 pt\hbox{$\mathchar \sim$}}
           \raise 1pt \hbox {$>$}}
\def\simlt{\rlap{\lower 3.5 pt\hbox{$\mathchar \sim$}}
           \raise 1pt \hbox {$<$}}
\newcommand{\DoneS}{{\Delta^{(3)}_{1}}}
\newcommand{\DtwoS}{{\Delta^{(3)}_{2}}}
\newcommand{\DfourS}{{\Delta^{(3)}_{4}}}
\newcommand{\DfiveS}{{\Delta^{(3)}_{5}}}
\newcommand{\DoneM}{{\Delta_{1}}}
\newcommand{\DtwoM}{{\Delta_{2}}}

\newcommand{\intS}{\int_{-\pi}^{\pi}\frac{d^{3}l}{(2\pi)^{3}}}
\newcommand{\intM}{\int_{-\pi}^{\pi}\frac{d^{4}l}{(2\pi)^{4}}}
\newcommand{\lm}{{\vec{l}}}
\begin{document}
\draft
\preprint{\vbox{\hbox{HUPD-9831}}}
\title{
$O(a\alpha_s)$ matching coefficients for the $\Delta B$=2 operators\\
in the lattice static theory
}
\author{
K-I. Ishikawa, T. Onogi and N. Yamada
}
\address{
Department of Physics, Hiroshima University,
Higashi-Hiroshima 739-8526, Japan
}
\maketitle

\begin{abstract}
We present the perturbative matching coefficient to $O(a\alpha_s)$
which relates the $\Delta B$=2 operator in the continuum
to that of the lattice static theory,
which is important in the accurate extraction
of the continuum value of the $B_B$ from lattice simulations.
The coefficients are obtained by the one-loop calculations in both
of the continuum and lattice theory.
We find that
two new dimension seven operators appear at the $O(a\alpha_s)$
with the $O(1)$ coefficients.
We also discuss the possible cancellation of $O(a\alpha_s)$ correction 
in the ratio
$B_B=\langle\overline{B}|\Op_L|B\rangle/
((8/3)(f_{B}M_{B})^2)$
qualitatively.
\end{abstract}

\section{Introduction}
\label{sec:introduction}
One of the most important issues in particle physics is
the origin of masses and CP violation.
CKM matrix elements are believed to play a key role to
probe the physics behind it.
Despite a lot of efforts in various approaches,
the matrix element $V_{td}$
which can be determined from $B^0-\overline{B^0}$ mixing
is still only poorly known
because of a lack of accuracy
in the involved hadronic matrix element. 
The hadronic matrix element is parameterized
using the $B$ meson decay constant $f_B$ and the bag parameter $B_B$
as $B_Bf_B^2$,
so it is quite crucial to compute them with high precision.
For this purpose,
the lattice QCD has been considered
to be one of the most reliable approaches.
So far most of the efforts have been devoted
to the $B$ meson decay constant.
At the early stage,
the decay constants in the static approximation and
from the extrapolation from light quarks were computed.
It was found that
both the lattice cutoff dependence and heavy quark mass dependence are
significantly large.
Later the scaling behavior
for the lattice spacing $a$~\cite{sofarfb,jlqcd_nrfb}
and the heavy quark mass $1/m_Q$~\cite{OUR}
have been investigated carefully and
the best estimate on $f_B$ from the quenched Lattice QCD is
now $f_B = 165(20)$ MeV~\cite{draper}.
On the other hand, until recently, the bag parameter has been
calculated only either in the static limit or
by the naive extrapolation from light quarks.
In this respect,
careful studies of systematic errors of the bag parameter are
still missing.

In general, in order to get a continuum result
of the physical quantity such as $f_B$ from lattice simulation,
we have to compute physical quantities on different lattices
and extrapolate the results to the continuum.
Therefore the final results have smaller errors
if the cutoff dependence is smaller.
It was found that
the $O(a)$ improvements of the action and lattice operators
in Symanzik approach significantly reduce
the lattice cutoff dependences of various matrix elements.
For the heavy-light axial vector current,
such kind of improvements have been accomplished
by Morningstar and Shigemitsu~\cite{MS} in the lattice NRQCD formalism.
They found that
the additional operator mixed at the $O(a\alpha_s)$ and
the inclusion of the effect significantly reduced
the value of $f_B$ at the finite lattice spacing and
it was also the case in the static limit.
In contrast to the decay constant,
the $O(a\alpha_s)$ mixing effect has not been studied for $B_B$.
One reason is that
only the operator matching of $O(\alpha_s)$ has been done
in Refs.~\cite{FHH,BP,PS,GR} so far.
Although previous simulations have not shown
a clear cutoff dependence of the $B_B$~\cite{GR,UKGM},
it would be very important
to study the $O(a\alpha_s)$ mixing effect explicitly 
in order to obtain the precise value of $B_B$.

The purpose of this paper is
to investigate the $O(a\alpha_s)$ effect for the $B_B$.
We perturbatively compute
the operator matching coefficients of
the static-clover $\Delta B$=2 operators up to $O(a\alpha_s)$.  
We use the notation defined by the authors in Refs.~\cite{FHH,BP}.

Phenomenologically important quantity might be
the product of $B_Bf_B^2$
which is just the expectation value of $\Delta B$=2 operator.
Therefore it seems sufficient
to improve only the $\Delta B$=2 operator.
To determine $f_B$ and $B_B$ separately, however,
would have some more advantage
from a technical point of view~\cite{draper}.
Since the $O(a\alpha_s)$ improvement for $B_B$ requires
the improvements of both of the heavy-light axial vector current and
the $\Delta B$=2 operator,
we also mention the result for the heavy-light current
for completeness.

The paper is organized as follows.
In sections \ref{sec:axial_vector} and \ref{sec:4fermi},
our main results,
the matching coefficients to the $O(a\alpha_s)$
for the heavy-light current and the $\Delta B$=2 operator,
are shown, respectively.
In section \ref{sec:discussion},
we discuss the impact of our results
on the determination of the $B_B$
using the typical values of the parameters involved and
the consistency with the previous observations
for the cutoff dependence of the $B_B$.
Finally we conclude in section \ref{sec:conclusion}.
The appendices are devoted to
some details in this calculation.

Throughout this paper,
we choose Feynman gauge ($\alpha$=1) and
the light quark mass $m_q$ is set to be zero.
The ultraviolet divergences appearing in the continuum calculation
are regulated by dimensional regularization
and 
the continuum operators are renormalized with $\overline{\rm MS}$ scheme,
while the infrared divergences are regulated by the gluon mass $\lambda$
in both of the continuum and lattice theory.
The operators with superscript ``con'' and ``lat'' 
define the continuum operators and the lattice operators,
 respectively.
In our convention, $\gamma_5$ always anticommutes with 
$\gamma_{\mu}$. We give all the equations in Euclidean form.

\section{Static heavy-light current}
\label{sec:axial_vector}
In this section, we present the matching coefficients of
the static-light current operators
which are relevant to the determinations of
the form factors of the static to light decays
as well as the following discussion.
Our lattice gauge action is the standard Wilson plaquette action.
For the light quark
we use the $O(a)$-improved SW quark action~\cite{SW}
with the clover coefficient $c_{\rm sw}$ and,
in contrast to Ref.\cite{BP},
we do not incorporate the rotation operator
associated with the clover fermion in the current operator.

In the following, we describe the lattice static quark.
In the static limit, the quark action is separated into two pieces
in Dirac basis, namely one for the static quark $b'$ and
the other for the static antiquark $\tilde{b'}$.
Both are two-component fields which are related to the relativistic
four-component field $b$ as
\be
b = \left( \begin{array}{c} b'\\ \tilde{b'}^\dag
           \end{array}
    \right),\ \
\bar{b} = \left( b'^\dag \  - \tilde{b'} \right) \label{spinor}.
\ee
In our convention, the action is given by
\be
      S^{\rm stat} 
&=&   \sum_{x,y}b'^{\dag i}_\alpha(x)
      \big[\  \delta_{x,y}\delta^{ij}
            - U_4^{\dag ij}(y)\delta_{x-\hat{4},y}
      \big] \delta_{\alpha\beta}
                b'^j_\beta(y) \no\\
& & + \sum_{x,y}\left(-\tilde{b'}^i_{\alpha'}(x)\right)
      \big[\  \delta_{x,y}\delta^{ij}
            - U_4^{ij}(x)\delta_{x+\hat{4},y}
      \big] \delta_{\alpha'\beta'}
                \tilde{b'}^{\dag j}_{\beta'}(y),
\ee
where $\alpha$ ($\alpha'$) and $\beta$ ($\beta'$)
run over 1 and 2 (3 and 4).
Our Feynman rules for the lattice static quark and antiquark
are obtained from the above action through the standard procedure.
The heavy quark (antiquark) propagates only forward (backward)
in time direction on the lattice.

To determine the matching coefficients upto $O(a\alpha_s)$,
(i) we calculate the heavy to light on-shell scattering amplitudes
through the following operator with arbitrary gamma matrix $\Gamma$,
\be
J_\Gamma^{(0)} = \overline{q} \Gamma b \no,
\ee
in the continuum full theory upto the one-loop order,
expand the resulting expression with respect to the momenta of
external quarks at their rest frame,
which is required to obtain the matching coefficients
through desired order $O(a\alpha_s)$, and
take the static limit of the heavy quarks.
(ii) We repeat the similar calculation to step (i)
on the lattice static theory.
(iii)Finally we express the continuum operators in terms of
the lattice operators with appropriate matching coefficients
which are adjusted to coincide the one-loop scattering amplitude
of both theories through $O(a\alpha_s)$.
In this matching procedure, we have two coupling constants,
$\alpha_s^{\rm \overline{MS}}$ in the continuum theory and
$\alpha_s^{\rm lat}$ on the lattice theory.
Through this paper, both coupling constants are rewritten
in terms of the $V$-scheme coupling~\cite{LM} at one-loop order.

According to step (i), we calculate the scattering amplitude
with an initial heavy quark carrying momentum $\p$ and
a final light quark carrying momentum $\k$.
The resulting expression is
\be
       \langle q(\k) |\ J_\Gamma^{(0)\rm con}\ | b(\p) \rangle
&=&    \Bigg[\ 1 + \frac{\alpha_s}{4\pi}C_F\ 
             \Bigg(
                      \left( \frac{1}{4}H^2 - \frac{5}{2} \right)
                      \ln\left(\frac{\mu^2}{m_b^2} \right)\no\\
& &                 - \frac{3}{2}\ln\left(\frac{\lambda^2}{\mu^2}
                                    \right)
                    - \frac{HG}{2} + \frac{3}{4}H^2 - HH'
                    - \frac{11}{4}
             \Bigg)
       \Bigg]
       \langle J_\Gamma^{(0)} \rangle_0 \no\\
& &  + \frac{\alpha_s}{4\pi}C_F\ G\ \frac{8\pi}{3a\lambda}
       \langle J_\Gamma^{(1)} \rangle_0 \label{curcon},
\ee
where the symbol $\langle\cdots\rangle_0$ denotes
the tree level expectation value
between the same initial and final states
as those of the left hand side, $C_F=(N_c^2-1)/2N_c$ with 
number of color $N_c$, $m_b$ is the heavy quark mass,
and $J_\Gamma^{(1)}\equiv\overline{q}(a\nabd\cdot\gam)\Gamma b$.
The renormalization scale for the amplitude is $\mu$.
The definitions of $H$, $G$ and $H'$ are the same as those in
Ref.~\cite{EH}.
In deriving Eq.~(\ref{curcon}),
we use the equation of motion for the light quark,
$\overline{q}\gamma_4 k_4=-\overline{q}\gam\cdot\k$
and also that for the heavy quark,
${\gamma}_4 u_b = u_b$, to simplify the result.

Repeating the similar calculation to the continuum theory
according to step (ii),
we obtain the corresponding amplitude on the lattice as follows.
\be
      \langle q(\k) |\ J_\Gamma^{(0)\rm lat}\ | b(\p) \rangle
&=&   \Bigg[ 1 + \frac{\alpha_s}{4\pi}C_F\
            \left(
                   - \frac{3}{2}\ln(a^2\lambda^2)
                   + A_\Gamma^{(0)} + A_\Gamma^{I(0)}
                   + \frac{1}{2}u_0^{(2)} 
            \right)
      \Bigg]\langle J_\Gamma^{(0)} \rangle_0 \no\\
& & + \frac{\alpha_s}{4\pi}C_F\
      \left(
               G\ \frac{8\pi}{3a\lambda}
             + r\ ( 1 - c_{\rm sw} )\ln(a^2\lambda^2)
             + A_\Gamma^{(1)} + A_\Gamma^{I(1)}
      \right)
      \langle J_\Gamma^{(1)} \rangle_0 \label{curlat},
\ee
where
\be
      A_\Gamma^{(0)}
&=&   d_1 +  d_2 G
    + \frac{1}{2}( e^{(R)} + f ), \\
      A_\Gamma^{I(0)}
&=&   - d^I G + \frac{1}{2} f^I, \\
      A_\Gamma^{(1)}
&=&   U G +  V , \\
      A_\Gamma^{I(1)}
&=&   U^I G + V^I.
\ee
The renormalization scale for the amplitude is $a^{-1}$.
$A_\Gamma^{(0)}$ and $A_\Gamma^{I(0)}$ correspond to
$A_\Gamma$ and $A_\Gamma^{I}$ in Ref.~\cite{BP}, respectively,
and the numerical values of $d_1$, $d_2$, $e^{(R)}$ and $f$
are tabulated in Refs.~\cite{FHH,BP,EH}.
Although our explicit form of the integrand of $d^I$
completely agrees with that of Ref.\cite{BP},
the numerical value of $d^I$
is slightly larger in magnitude than that of Ref.~\cite{BP},
and the value is tabulated in Table \ref{tab:PERT}.
$U$, $U^I$, $V$ and $V^I$ are new contributions at $O(a\alpha_s)$.
Their explicit forms of the integrands are shown
in Appendix \ref{sec:appendix} and their numerical values
are tabulated in Table \ref{tab:PERT}.
The coefficients with the superscript $I$ vanish
when Wilson light quark is used~($c_{\rm sw}$=0),
which is the same notation as Ref.\cite{BP}.
$u_0^{(2)}$ comes from the tadpole improvement of
the light quark wave function renormalization,
for details see Appendix \ref{sec:wave_renorm}.

In step (iii), matching Eq.~(\ref{curcon}) to Eq.~(\ref{curlat}),
we obtain the following relation between the operators
in the continuum and lattice theory,
\be
      J_\Gamma^{(0)\rm con}
&=&   \Bigg[ 1 + \frac{\alpha_s}{4\pi}C_F\ \zeta_\Gamma^{(0)}\
      \Bigg]\ J_\Gamma^{(0)\rm lat}
    + \frac{\alpha_s}{4\pi}C_F\ \zeta_\Gamma^{(1)}\
      J_\Gamma^{(1)\rm lat} \no\\
&\equiv&   Z_\Gamma^{(0)}\ J_\Gamma^{(0)\rm lat}
         + Z_\Gamma^{(1)}\ J_\Gamma^{(1)\rm lat}
 \label{hlcurr},
\ee
where
\be
    \zeta_\Gamma^{(0)}
&=&   \left( \frac{1}{4}H^2 - \frac{5}{2} \right)
      \ln\left(\frac{\mu^2}{m_b^2} \right)
    - \frac{3}{2}\ln\left(\frac{\lambda^2}{\mu^2}\right)
    - \frac{HG}{2} + \frac{3}{4}H^2 - HH' - \frac{11}{4} \no\\
& & + \frac{3}{2}\ln(a^2\lambda^2)
    - A_\Gamma^{(0)} - A_\Gamma^{I(0)} - \frac{1}{2}u_0^{(2)} 
\label{eq:zeta_zero_Gamma}
  \\
    \zeta_\Gamma^{(1)}
&=& - r\ ( 1 - c_{\rm sw} )\ln(a^2\lambda^2)
    - A_\Gamma^{(1)} - A_\Gamma^{I(1)} 
\label{eq:zeta_one_Gamma}
\ee
Eq.~(\ref{eq:zeta_zero_Gamma}) has been calculated in
Refs.~\cite{FHH,BP} except for the differences of 
our inclusion of tadpole improvements and the wave function 
renormalization of lattice static quarks.
Eq.~(\ref{eq:zeta_one_Gamma}) is new result for the 
arbitrary static-light current.
For axial vector current and vector current the matching
coefficient for $J_\Gamma^{(1)\rm lat}$ has been calculated
with NRQCD action for heavy quarks in Ref.~\cite{MS}.
From Eq.~(\ref{hlcurr}) we observe that
the $O(a)$ operator $J_\Gamma^{(1)\rm lat}$ appears at this order,
which is considered to be a lattice artifact.
It is noted that
there is no linear divergence proportional to $1/\lambda$
in the coefficients,
while there is a logarithmic divergence unless $c_{\rm sw}=1$.
In the use of Wilson light quark~($c_{\rm sw}$=0), therefore,
we cannot match these operators consistently
due to this infrared mismatch
as previously pointed out in Refs.~\cite{MS,BP}.

The results of $\zeta_\Gamma^{(0)}$ and $\zeta_\Gamma^{(1)}$
for each $\Gamma$ are summarized in Table \ref{tab:hhd},
where $r=c_{\rm sw}=1$ and
the tadpole improvement is performed
by using the perturbative expression of
the critical hopping parameter.
The numerical values of $O(a\alpha_s)$ correction for
axial vector current and vector current are consistent
with those in Ref.~\cite{MS}\footnote{
Note that since there are some differences of
the definitions of the lattice operators and 
the matching coefficients between ours and theirs in Ref.~\cite{MS},
one would need to redefine our definitions to compare 
the results with theirs.}.
It should be noted that
the coefficient of $J_\Gamma^{(1)\rm lat}$ depends only on $G$,
$G=-1$ might lead to a large mixing effect,
while $G=1$ does not.
Actually the mixing effect leads to the significant change
for the $f_B$, which has been seen in Refs.~\cite{jlqcd_nrfb,SGO}.

\section{$\Delta B$=2 operator}
\label{sec:4fermi}
In this section,
we discuss the matching of the $\Delta B$=2 operator.
The matching procedure is completely common as
that for heavy-light current previously shown, and
we follow the previous step.
Before proceeding step (i),
we give the definitions of the operators.
\be
\Op_L &=& \left[ \overline{b}\gamma_\mu P_L q\right]
        \left[ \overline{b}\gamma_\mu P_L q\right],\no\\
\Op_S &=& \left[ \overline{b} P_L q\right]
        \left[ \overline{b} P_L q\right],\no\\
\Op_R &=& \left[ \overline{b}\gamma_\mu P_R q\right]
        \left[ \overline{b}\gamma_\mu P_R q\right],\no\\
\Op_N &=&   2 \left[ \overline{b}\gamma_\mu P_L q\right]
            \left[ \overline{b}\gamma_\mu P_R q\right]
          + 4 \left[ \overline{b} P_L q\right]
            \left[ \overline{b} P_R q\right],
            \no\\
\Op_{LD} &=& \left[ \overline{b}\gamma_\mu P_L q\right]
          \left[ \overline{b}\gamma_\mu P_L
                 (a \D\cdot\gam )q\right],\no\\
\Op_{ND} &=&   2 \left[ \overline{b}\gamma_\mu P_L q\right]
              \left[ \overline{b}\gamma_\mu P_R
                     (a \D\cdot\gam )q
              \right]
            + 4 \left[ \overline{b} P_L q\right]
              \left[ \overline{b} P_R
                   (a \D\cdot\gam )q\right], \no
\ee
where $P_L$=$1-\gamma_5$ and $P_R$=$1+\gamma_5$.

According to step (i), 
we calculate the two-body scattering 
amplitude for $\Op_L$ 
between the initial state 
with a heavy antiquark and a light quark
and the final state
with a heavy quark and a light antiquark
in the continuum theory.
The initial heavy antiquark carries momentum $\p_2$,
the initial light quark $\k_2$,
the final heavy quark $\p_1$, and
the final light antiquark $\k_1$.
We obtain the scattering amplitude in the continuum theory
at one-loop as
\be
& &   \langle \overline{q}(\k_1),b(\p_1) | \ \Op_L^{\rm con}\
      | q(\k_2),\overline{b}(\p_2) \rangle
 =    Z_q^{\rm con}Z_b^{\rm con}
      \sum_i\V_{\rm con}^{(i)}(\k_1,\p_1,\k_2,\p_2) \no\\
&=&   \left[
      1 + \frac{\alpha_s}{4\pi}\
          \left(
               2 \ln\left( \frac{m_b^2}{\mu^2} \right)
             - 4 \ln\left( \frac{\lambda^2}{m_b^2} \right)
             + C_L + \frac{7}{3}
          \right)
      \right] \langle \Op_L \rangle_0 \no\\
& & + \frac{\alpha_s}{4\pi}\ C_S
      \langle \Op_S \rangle_0
    + \frac{\alpha_s}{4 \pi}\ \frac{16\pi}{3a\lambda}
      \langle \Op_{ND} \rangle_0 \label{4fconef},
\ee
where the $\V_{\rm con}^{(i)}$ ($i$ runs over a-d)
denotes the contribution from each diagram 
in the continuum theory, which appear 
in Appendix \ref{sec:appendixc}.
The constants $C_L=-14$ and $C_S=-8$ appear in Refs.~\cite{FHH,BP}.

According  to step (ii),
we calculate the corresponding amplitude with the lattice theory
and obtain the result as follows.
\be
& & \langle \overline{q}(\k_1),b(\p_1) | \ \Op_L^{\rm lat}\
    | q(\k_2),\overline{b}(\p_2) \rangle
 = Z_q^{\rm lat}Z_b^{\rm lat}
   \sum_i\V_{\rm lat}^{(i)}(\k_1,\p_1,\k_2,\p_2) \no\\
&=& \Bigg[
      1 + \frac{\alpha_s}{4\pi}\
           \left( - 4 \ln( a^2\lambda^2 ) - D_L - D_L^I + \frac{7}{3}
                  + \frac{4}{3} u_0^{(2)}
           \right)
    \Bigg] \langle \Op_L \rangle_0 \no\\
&&+ \frac{\alpha_s}{4\pi}\ (- D_N - D_N^I )
    \langle \Op_N \rangle_0
  + \frac{\alpha_s}{4\pi}\ (- D_R - D_R^I )
    \langle \Op_R \rangle_0 \no\\
&&+ \frac{\alpha_s}{4\pi}\
           \left( - \frac{10}{3}\
                    r ( 1 - c_{\rm sw} )\ln(a^2\lambda^2)
                  - D_{LD} - D_{LD}^I
           \right)
    \langle \Op_{LD} \rangle_0 \no\\
&&+ \frac{\alpha_s}{4\pi}\ 
           \left(
                    \frac{16\pi}{3a\lambda}
                  - D_{ND} - D_{ND}^I
           \right)
    \langle \Op_{ND} \rangle_0\label{4fconlat},
\ee
where
\be
      D_L
&=& - \frac{10}{3} d_1 - \frac{1}{3} c - \frac{1}{3} v 
    - \frac{4}{3}( e^{(R)} + f ) + \frac{7}{3},\\
      D_L^I
&=& - \frac{1}{3} v^I - \frac{4}{3} f^I,\\
      D_N
&=&   2\ d_2,\\
      D_N^I
&=& - 2\ d^I,\\
      D_R
&=&   \frac{4}{3} w,\\
      D_R^I
&=&   \frac{4}{3} w^I,\\
      D_{LD}
&=&   \frac{10}{3} V,\\
      D_{LD}^I
&=&   \frac{10}{3} V^I,\\
      D_{ND}
&=& - 2 U,\\
      D_{ND}^I
&=& - 2 U^I.
\ee
The coefficients
$D_{L}$, $D^{I}_{L}$, $D_{N}$, $D^{I}_{N}$, $D_{R}$, and $D^{I}_{R}$
have been calculated in Refs.\cite{FHH,BP,PS,GR} and we use the same
notation as those in Refs.\cite{FHH,BP} for convenience.
The coefficients $D_{LD}$, $D^{I}_{LD}$,  $D_{ND}$, and
$D^{I}_{ND}$ are novel results of this paper.
$\V_{\rm lat}^{(i)}$ ($i$ runs over a-d) is the contribution from
each diagram in the lattice theory, which are shown in 
the Appendix \ref{sec:appendixc}.

According to step (iii), using Eqs.~(\ref{4fconef}) and
(\ref{4fconlat}) we match the the lattice operator and 
continuum one to $O(a\alpha_s)$. The obtained operator
identity is
\be
      \Op^{\rm con}_L
&=&   \sum_{X}
      Z_X\ \Op^{\rm lat}_X\label{4fmat},
\ee
where $X$ runs over \{$L,S,N,R,LD,ND$\},
\be
      Z_L
&=&   1 + \frac{\alpha_s}{4\pi}
          \Bigg(   6 \ln(a^2m_b^2) - 2 \ln(a^2\mu^2)
                + C_L + D_L + D_L^I - \frac{4}{3} u_0^{(2)}
          \Bigg), \label{i}\\
      Z_S
&=&   \frac{\alpha_s}{4\pi}\ C_S,\\
      Z_N
&=&   \frac{\alpha_s}{4\pi}\ \left( D_N + D_N^I \right),\\
      Z_R
&=&   \frac{\alpha_s}{4\pi}\ \left(  D_R + D_R^I \right),\\
      Z_{LD}
&=&   \frac{\alpha_s}{4\pi}\
      \left(\  \frac{10}{3} r\ ( 1 - c_{\rm sw} )\ln(a^2\lambda^2)
            + D_{LD} + D_{LD}^I \right),\\
      Z_{ND}
&=&   \frac{\alpha_s}{4\pi}\ \left(  D_{ND} + D_{ND}^I \right)\label{f}.
\ee
Here we omit the explicit arguments of $\mu$ and $a^{-1}$,
which are introduced by the renormalization procedure,
for the operators $O_{X}$ and the coefficients $Z_{X}$ 
without ambiguity.
We find that the above results to $O(\alpha_s)$ agrees with those of
Refs.~\cite{FHH,BP}
except for the coefficient $D_R^I$
in Ref.~\cite{BP} (see Appendix \ref{sec:appendixc}).
The correct value of $D_R^I$ including double rotation operator
has been already obtained in Refs.\cite{PS,GR}
and our $D_R^I$ is consistent with them.
Two new operators $\Op^{\rm lat}_{LD}$ and $\Op^{\rm lat}_{ND}$
mix at the $O(a\alpha_s)$.
It should be noted that
the coefficients of
the new operators have completely common integrands to
those of $J_\Gamma^{(1)\rm lat}$ in the heavy-light current.
The use of Wilson light quark~($c_{\rm sw}$=0) leads to
the mismatch of infrared behavior
between continuum and lattice theory
as in the case of heavy-light current.

When $c_{\rm sw} = r = 1$ and
the tadpole improvement by the critical hopping parameter are chosen 
for the numerical estimate of Eqs.~(\ref{i})-(\ref{f}),
Eq.~(\ref{4fmat}) becomes
\be
    \Op_L^{\rm con}
&=& \Bigg[\ 1 + \frac{\alpha_s}{4\pi}
           \Bigg(   6 \ln(a^2m_b^2) - 2 \ln(a^2\mu^2)
                   - 35.15 \Bigg)
  \Bigg] \Op_L^{\rm lat} \no\\
&& + \frac{\alpha_s}{4\pi}\ (-8)\ \Op_S^{\rm lat}
   + \frac{\alpha_s}{4\pi}\ \left( -6.16 \right)
     \Op_N^{\rm lat}
   + \frac{\alpha_s}{4\pi}\ \left(  -0.52 \right)
     \Op_R^{\rm lat} \no\\
&& + \frac{\alpha_s}{4\pi}\
     \left( - 17.20 \right)
     \Op^{\rm lat}_{LD}
   + \frac{\alpha_s}{4\pi}\ \left( - 9.20 \right)
     \Op^{\rm lat}_{ND} \label{numerical}.
\ee
It is found that
the coefficients of two new operators are
17.20/4$\pi$ and 9.20/4$\pi$, respectively, and are of $O(1)$.
This means the possibility of large $O(a\alpha_s)$ correction
for $\Op_L^{\rm con}$ as in the case of axial vector current,
though the lattice matrix elements of $\Op_{LD}^{\rm lat}$
and $\Op_{ND}^{\rm lat}$ are not yet known.

\section{Discussion}
\label{sec:discussion}
In the previous section,
we pointed out that
the $\Delta B=2$ operator might receive
a large $O(a\alpha_s)$ correction.
For the rigorous investigation of the $O(a\alpha_s)$ effect,
we must rely on the future works.
On the other hand,
the previous simulations have not shown
a clear cutoff dependence of the $B_B$ and
seem to imply that
the vacuum saturate approximation (VSA) is plausible within 10\% level
around the used lattice cutoff
scale ($\sim$ 2-3 GeV)~\cite{GR,UKGM,CDM,YA}.
In this section, therefore,
we attempt assuming the VSA for the lattice matrix elements to
estimate the $O(a\alpha_s)$ effects for the $B_Bf_B^2$ and $B_B$
using the results of the previous sections and then
investigate the consistency of our result
with the previous simulations.
Although this analysis is quite rough,
we believe that
it is possible to find some, at least, qualitative features.

Let us discuss the $O(a\alpha_s)$ correction for
$\langle \overline{B^0}|\ \Op_L^{\rm con}\ | B^0\rangle$
using the VSA.
Under the VSA, 
the relevant lattice matrix elements take the following values,
\be
& &   \langle \overline{B^0}|\ \Op_L^{\rm lat}\
      | B^0\rangle^{\rm (VSA)} 
 =    \langle \overline{B^0}|\ \Op_R^{\rm lat}\
      | B^0\rangle^{\rm (VSA)} \no\\
&=&   \langle \overline{B^0}|\ \Op_N^{\rm lat}\
      | B^0\rangle^{\rm (VSA)}
 =  - \frac{8}{5}
      \langle \overline{B^0}|\ \Op_S^{\rm lat}\
      | B^0\rangle^{\rm (VSA)}
 =    \frac{8}{3}\left(f_B^{(0)\rm lat} M_B \right)^2,\label{vsa1}\\
& &   \langle \overline{B^0}|\ \Op_{LD}^{\rm lat}\
      | B^0\rangle^{\rm (VSA)}
 =    \langle \overline{B^0}|\ \Op_{ND}^{\rm lat}
      | B^0\rangle^{\rm (VSA)}
 =  - \delta f_B^{\rm lat}
      \frac{8}{3}\left(f_B^{(0)\rm lat} M_B \right)^2, \label{vsa2}
\ee
where 
$f_B^{(0)\rm lat}M_B \equiv
\langle 0 | J_{\gamma_5\gamma_4}^{(0)\rm lat} | \overline{B^0}\rangle$
 and
$\delta f_B^{\rm lat} \equiv
 \langle 0|J_{\gamma_5\gamma_4}^{(1)\rm lat}|\overline{B^0}\rangle/
 \langle 0|J_{\gamma_5\gamma_4}^{(0)\rm lat}|\overline{B^0}\rangle$.
Substituting Eqs.~(\ref{vsa1}) and (\ref{vsa2}) into
Eq.~(\ref{numerical}), we obtain 
\be
&& \hspace{-3em}
 \langle \overline{B^0}|\ \Op_L^{\rm con}\ | B^0\rangle
\vsa 
 \langle \overline{B^0}|\ \Op_L^{\rm con}\ | B^0\rangle^{\rm (VSA)}
\no\\
&=&
     \frac{8}{3}\left(f_B^{(0)\rm lat} M_B \right)^2
     \Bigg[\ 1 + \frac{\alpha_s}{4\pi}
     \Bigg(    6 \ln(a^2m_b^2) - 2 \ln(a^2\mu^2)
           - 36.83
           + 26.40\ \delta f_B^{\rm lat}
     \Bigg)\
     \Bigg],
\ee
where the last term with $\delta f_B^{\rm lat}$
is essentially due to the $O(a\alpha_s)$ effect.
We can use the data calculated by Ali Khan {\it et al.} in Ref.~\cite{SGO} 
to guess the value of $\delta f_B^{\rm lat}$ in the static limit.
In our estimate, their finite mass results at $\beta=6.0$ imply
$\delta f_B^{\rm lat}\sim-0.5$ in the static limit.
Using the coupling constants at the corresponding lattice
with Lepage and Mackenzie prescription~\cite{LM},
$\alpha_s\sim 0.15$-$0.25$,
we find that
the magnitude of the $O(a\alpha_s)$ correction 
for
$\langle \overline{B^0}|\ \Op_L^{\rm con}\ | B^0\rangle^{\rm (VSA)}$
is very large, about 15-25\%.
Although this analysis is naive estimate of
$O(a\alpha_s)$ correction using the VSA, this suggests that there are 
large contribution from $O(a\alpha_s)$ correction for
$\Op_L^{\rm con}$ and the improvement of $O(a\alpha_s)$ 
should be necessarily included.

Now we turn to the $B_B$.
The $B_B$ is defined by
\be
    B_B
&=& \frac{\langle \overline{B^0}|\ \Op^{\rm con}_L\ | B^0\rangle}
         {\frac{8}{3}\left( f_B M_B \right)^2 }. \label{defB}
\ee
To improve the $B_B$ in consistent way,
we should include the $O(a\alpha_s)$ improvements
of both of the numerator and denominator of Eq.~(\ref{defB}).
Substituting Eqs.~(\ref{hlcurr}) and (\ref{4fmat})
into Eq.~(\ref{defB}) and
linearizing the resulting expression in $\alpha_s$
according to the discussion of Ref.~\cite{CDM},
we obtain the $B_B$ as follows.
\be
    B_B
&=& \sum_X \omega_X B_X^{\rm lat}
  - 2\ \omega_1\delta f_B^{\rm lat} B_L^{\rm lat}, \no
\ee
where $X$ runs over $\{L,S,N,R,LD,ND\}$,
\be
      \omega_X
&=&   \frac{Z_X}{(Z_{\gamma_5\gamma_4}^{(0)})^2}, \no\\
      \omega_1
&=&   \frac{Z_{\gamma_5\gamma_4}^{(1)}}
           {Z_{\gamma_5\gamma_4}^{(0)}},\no\\
      B_X^{\rm lat}
&=&   \frac{\langle \overline{B^0}|\
            \Op_X^{\rm lat}\
            | B^0 \rangle} 
           {\frac{8}{3}
            \left( f_B^{(0)\rm lat}M_B \right)^2}.\no
\ee
In the VSA, using Eqs.~(\ref{vsa1}) and (\ref{vsa2})
we obtain the following expression for $B_B$ to
$O(a\alpha_s)$.
\be
 B_B \vsa B_B^{\rm (VSA)} 
&=&
       \left( \omega_L + \omega_R + \omega_N - \frac{5}{8} \omega_S
       \right)
     - \left( \omega_{LD} + \omega_{ND} + 2\ \omega_1
       \right) \delta f_B^{\rm lat}, \no \\
&=& \left( 1 + \frac{\alpha_s}{4\pi} D \right) 
  - \left(     \frac{\alpha_s}{4\pi} E \right) \delta f_B^{\rm lat},
\ee
where the term with $\delta f_B^{\rm lat}$ comes from
the $O(a\alpha_s)$ improvements again.
The coefficients $D$ and $E$ are given as follows.
\be
D &=& \Bigg[  2 \ln\left(\frac{m_b^2}{\mu^2}\right)
               - \frac{14}{3}
               - \frac{2}{3}\left( d_1 + d_2 - d^I \right)
               - \frac{1}{3} c
               - \frac{1}{3}( v + v^I )
               + \frac{4}{3}\ ( w + w^I )
      \Bigg],\label{coefinbb1}\\
E &=& \frac{2}{3}\
    \Bigg[  r\ (1-c_{\rm sw}) \ln(a^2\lambda^2)
           + U + U^I + V + V^I 
    \Bigg]. \label{coefinbb2}
\ee
In deriving Eqs.~(\ref{coefinbb1}) and (\ref{coefinbb2}),
there are some cancellations between the coefficients of
the $\Delta B =2$ operator and the axial vector current.

Now let us roughly estimate the $O(a\alpha_s)$ effect in the 
$B_B^{\rm (VSA)}$ numerically.
When $r = c_{\rm sw}=1$ is chosen, we obtain
$D = 2 \ln(m_b^2/\mu^2) -3.72$ and $E=-0.37$.
Using the data of $\delta f_B^{\rm lat}$ and
the coupling constants as before,
we find that the $O(a\alpha_s)$ effect for 
the $B_B^{\rm(VSA)}$ is smaller than 1\%.
Of course such a drastic cancellation would not
take place in reality due to deviations from VSA,
but at least the present analysis suggests that
there is a possibility of significant cancellation of
$O(a\alpha_s)$ corrections in $B_B$.
This is consistent with the observation from the previous simulations
that there is not clear cutoff dependence in $B_B$.

\section{conclusion}
\label{sec:conclusion}
In this paper,
we reported the coefficients of the $O(a)$ operators
which are newly induced at the $O(a\alpha_s)$
in the perturbative continuum-lattice operator matching
of the heavy-light current and $\Delta B=2$ operator.
We also roughly estimated the $O(a\alpha_s)$ effect
on the $B_Bf_B^2$ and the $B_B$
using the VSA in the lattice hadronic matrix elements.
Although the $O(a\alpha_s)$ effect is
significant in the determinations of $f_B$ and $B_Bf_B^2$,
it seems that the effect is not so for $B_B$, at least,
in this VSA analysis
because the cancellation between the $O(a\alpha_s)$ effects
in the numerator and denominator
does work well.
Therefore the previous works,
which imply that there is no cutoff dependence in $B_B$,
seem to be consistent with our analysis.
Now, however, that
the $O(a\alpha_s)$ improvement for the $f_B$
has been already done,
in order to calculate the $B_B$ in a consistent way 
the $O(a\alpha_s)$ operators should be included in the calculation.
For the precise determination,
it is also required to include the finite mass correction into
both calculations of the matrix element and the matching coefficients.

\section*{Acknowledgment}
We would like to thank S.~Hashimoto for useful discussion.
T.~O. and N.~Y would also like to thank Summer Institute 98
at Kyoto supported by the Grant-in Aid
for Scientific Research on Priority Areas (Physics of CP Violation)
from the Ministry of Education, Science and Culture of Japan,
where this work has started.

T.~O. is supported by the Grants-in-Aid of the Ministry 
of Education~(No.10740125).

\appendix
\section{}
\label{sec:wave_renorm}
Here we show the wave function renormalization constants
for each external quark line in each theory,
\be
    Z_q^{\rm con}
&=& 1 - \frac{\alpha_s}{4\pi}C_F
    \left[   {\cal A} - \ln\left(\frac{\lambda^2}{\mu^2}\right)
          - \frac{1}{2} \right] \no,\\
    Z_b^{\rm con}
&=& 1 - \frac{\alpha_s}{4\pi}C_F
    \left[   {\cal A} - \ln\left(\frac{m_b^2}{\mu^2}\right)
          - 2 \ln\left(\frac{m_b^2}{\lambda^2}\right) + 4 \right]
     \no,\\
    Z_q^{\rm lat}
&=& 1 + \frac{\alpha_s}{4\pi}C_F
        \left[\  \ln(a^2\lambda^2) + f + f^I + u_0^{(2)}
        \right] \no,\\
    Z_b^{\rm lat}
&\equiv& Z_Q^{\rm lat}
 =  Z_{\chi}^{\rm lat}
 =  1 + \frac{\alpha_s}{4\pi}C_F
        \left[ - 2 \ln(a^2\lambda^2) + e^{(R)}
        \right], \no
\ee
where $f$, $f^I$ and $e^{(R)}$
can be found in Refs.~\cite{FHH,BP}.
In above equations
${\cal A}=1/\epsilon + \ln(4\pi) - \gamma_E$ and
$u_0^{(2)}$ is perturbative coefficient of the
tadpole improvement factor defined by
$u_0=1+\alpha_s C_F u_0^{(2)}$.
The coefficient $u_0^{(2)}$ is obtained through the calculation
of the mean plaquette value or the critical hopping parameter,
$u_0^{(2)}=  -\pi^2$ or 
$u_0^{(2)}=  -\left[   4.4259 + 8.4327\ r
                     - 4.8619\ c_{\rm sw} \right]$, respectively.

\section{}
\label{sec:appendix}
Here we show the explicit forms of the integrands
for $U$, $U^I$, $V$ and $V^I$,
which first appear in the $O(a\alpha_s)$.
For shorthand notation, we define the following quantities,
\be
\DoneM  &=& \sum_{\mu=1}^4\sin^2\left(\frac{l_\mu}{2}\right), \no\\
\DtwoM  &=& \sum_{\mu=1}^4\sin^2(l_\mu) + 4r^2(\DoneM)^2, \no\\
\DoneS  &=& \sum_{\mu=1}^3\sin^2\left(\frac{l_\mu}{2}\right),\no\\
\DtwoS  &=& \sum_{\mu=1}^3\sin^2(l_\mu) + 4r^2(\DoneS)^2, \no\\
\DfourS &=& \sum_{\mu=1}^3\sin^2(l_\mu), \no\\
\DfiveS &=& \sum_{\mu=1}^3\sin^2(l_\mu)
                          \sin^2\left(\frac{l_\mu}{2}\right). \no
\ee
Using the above convention,
\be
      U
&=&   (4\pi)^{2} \intS \left[
      \frac{1}{12\DoneS\DtwoS}\left(3+(3r^{2}-1)\DoneS\right)
      \right.\no \\
& &   \left.
    - \frac{1}{12\DoneS(\DtwoS)^{2}}
      (\DfourS - 2\DfiveS + 2r^{2}\DoneS\DfourS)
    - \frac{2}{3(\lm^{2})^{2}}\theta(1-\lm^{2}) \right]-\frac{16}{3},
\no \\
      U^{I}
&=&   (4\pi)^{2} r^{2} \intS\left[
      \frac{\DfourS}{48\DoneS\DtwoS}
    - \frac{1}{12(\DtwoS)^{2}}
      (\DfourS - 2\DfiveS + 2r^{2}\DoneS\DfourS)\right],
\no \\
      V
&=&   (4\pi)^{2} r \intM\left[
    - \frac{1}{4\DtwoM} \right.\no \\
& & - \frac{1}{12\DoneM(\DtwoM)^{2}}
      \left\{ 12\left(1+2\DoneS+2(r^{2}-1)\DoneM\right)
      (1-\DoneM+\DoneS)\DoneS\right.\no\\ 
& &   \left.\left.
   + (\DfourS-2\DfiveS+2r^{2}\DfourS\DoneM)\right\}
   + \frac{1}{(l^{2})^{2}}\theta(1-l^{2})\right],
\no\\
     V^{I}
&=&  (4\pi)^{2} r \intM\left[
     \frac{1}{12\DoneM(\DtwoM)^{2}}\left\{
     \left(1+2\DoneS+2(r^2-1)\DoneM\right)\DfourS \right. \right.\no\\
& &  \left.\left.
   + (\DfourS-2\DfiveS+2r^{2}\DfourS\DoneM)\right\}(1-\DoneM+\DoneS)
   - \frac{1}{l^{2}}\theta(1-l^{2})\right].\no
\ee

\section{}
\label{sec:appendixc}
Here we show the contribution from each diagram explicitly.
In the continuum, the each contribution is as follows.
\be
\V_{\rm con}^{(a)}
&=&   \langle \Op_L \rangle_0, \no\\
\V_{\rm con}^{(b)} 
&=&   \frac{\alpha_s}{4 \pi}
      \left[  \frac{10}{3} {\cal A} 
            - \frac{10}{3}\ln\left(\frac{\lambda^2}{\mu^2}\right)
            - \frac{11}{3}
      \right]
      \langle \Op_L \rangle_0
    - \frac{\alpha_s}{4 \pi}\ 8\ \langle \Op_S \rangle_0
    + \frac{\alpha_s}{4 \pi}\ \frac{16\pi}{3a\lambda}
      \ \langle \Op_{ND} \rangle_0, \no\\
\V_{\rm con}^{(c)}
&=&   \frac{\alpha_s}{4 \pi}
     \left[
        -\frac{4}{3}{\cal A}
        + \frac{4}{3}\ln\left(\frac{m_b^2}{\mu^2}\right)
        - \frac{2}{3}\ln\left(\frac{\lambda^2}{m_b^2}\right)
        - \frac{5}{3}
     \right]
     \langle \Op_L \rangle_0, \no\\
\V_{\rm con}^{(d)}
&=&   \frac{\alpha_s}{4 \pi}
      \left[
         - \frac{4}{3}{\cal A}
         + \frac{4}{3}\ln\left(\frac{\lambda^2}{\mu^2}\right)
         - \frac{5}{3}
      \right]
      \langle \Op_L \rangle_0, \no
\ee
where (a) corresponds the tree diagram,
(b) those with the gluon
connecting the static and the light quarks, 
(c) those connecting the static quark and the static antiquark,
and (d) those connecting the light quark and the light antiquark.
And on the lattice,
\be
\V_{\rm lat}^{(a)}
&=&   \langle \Op_L \rangle_0, \no\\
\V_{\rm lat}^{(b)}
&=&   \frac{\alpha_s}{4 \pi}\ \frac{10}{3}\
      \left[ - \ln(a^2\lambda^2) + d_1 \right]
      \langle \Op_L \rangle_0 \no\\
&&  + \frac{\alpha_s}{4 \pi}\ 2\
      \left[ - d_2 + d^I \right]\no
      \langle \Op_N \rangle_0\no\\
&&  + \frac{\alpha_s}{4 \pi}\ \frac{10}{3}
      \left[\  r ( c_{\rm sw}- 1 )\ln(a^2\lambda^2)
             - ( V + V^I ) \right]
      \langle \Op_{LD} \rangle_0\no\\
&&  + \frac{\alpha_s}{4 \pi}\ 2
      \left[\ \frac{8\pi}{3a\lambda} + ( U + U^I )\right]
      \langle \Op_{ND} \rangle_0, \no\\
\V_{\rm lat}^{(c)}
&=&   \frac{\alpha_s}{4 \pi}\ \frac{1}{3}\
      \left[ -\ 2 \ln(a^2\lambda^2) + c \right]
      \langle \Op_L \rangle_0, \no\\
\V_{\rm lat}^{(d)}
&=&   \frac{\alpha_s}{4 \pi}\ \frac{1}{3}\
      \left[\ 4 \ln(a^2\lambda^2) + ( v + v^I ) \right]
      \langle \Op_L \rangle_0 \no\\
& & + \frac{\alpha_s}{4 \pi}\ \frac{4}{3}\
      \left[- ( w + w^I ) \right]
      \langle \Op_R \rangle_0. \label{w}
\ee
The calculation is straightforward though slightly lengthy.
The full use of the equations of motion
for the heavy and the light quarks and
of the identities for $\gamma$ matrices
sometimes leads to simplification,
in particular for the derivation of $\V_{\rm lat}^{(d)}$.
We find that
our result of $\V_{\rm lat}^{(d)}$ is inconsistent with
Eqs.~(B.16) and (B.25) of Ref.~\cite{BP} provided
the sign of the numerical values tabulated in T{\scriptsize ABLE} 3 of 
the reference was correct,
which has been already pointed out in Refs.\cite{PS,GR}.


\begin{table}[h]
\begin{center}
\caption{The numerical values of $d^I$, $U$, $U^I$, $V$ and $V^I$
for each value of $r$.}
\label{tab:PERT}
\begin{tabular}{cccccc}
  $r$  & 1.00 & 0.75 & 0.50 & 0.25 & 0  \\
$d^{I}$&-4.14 &-3.74 &-3.12 &-2.04 & 0  \\
  $U$  & 4.89 & 5.27 & 6.16 & 8.26 &12.72 \\
$U^{I}$&-0.29 &-0.11 & 0.02 & 0.06 & 0  \\
  $V$  &-7.14 &-7.51 &-7.72 &-6.99 & 0  \\
$V^{I}$& 1.98 & 1.82 & 1.51 & 0.98 & 0 \\
\end{tabular}
\end{center}
\end{table}
\begin{table}[h]
\begin{center}
\caption{The results of the heavy-light current matching and
$H$, $H'$ and $G$ for each $\Gamma$.}
\label{tab:hhd}
\begin{tabular}{cccccc}
$\Gamma$           &  $H$ & $H'$ &  $G$
                   & $\zeta_\Gamma^{(0)}$ & $\zeta_\Gamma^{(1)}$ \\
$1$                &  4 &  1 &  1 
                   &  3 $\frac{3}{2}\ln(\mu^2/m_b^2)$
                    + $\frac{3}{2}\ln(a^2 m_b^2)$ - 2.25
                   &  0.56 \\
$\gamma_5$         & -4 & -1 & -1
                   &  3 $\frac{3}{2}\ln(\mu^2/m_b^2)$
                    + $\frac{3}{2}\ln(a^2 m_b^2)$ - 8.41
                   &  9.76 \\
$\gamma_i$         & -2 & -1 & -1
                   &  $\frac{3}{2}\ln(a^2 m_b^2)$ - 14.41
                   &  9.76 \\
$\gamma_4$         & -2 & -1 &  1
                   &  $\frac{3}{2}\ln(a^2 m_b^2)$ - 6.25
                   &  0.56 \\
$\gamma_5\gamma_i$ &  2 &  1 &  1
                   &  $\frac{3}{2}\ln(a^2 m_b^2)$ - 8.25
                   &  0.56 \\
$\gamma_5\gamma_4$ &  2 &  1 & -1
                   &  $\frac{3}{2}\ln(a^2 m_b^2)$ - 12.41
                   &  9.76 \\
$\sigma_{4i}$      &  0 &  1 & -1
                   & - $\frac{3}{2}\ln(\mu^2/m_b^2)$
                     + $\frac{3}{2}\ln(a^2 m_b^2)$ - 14.41
                   &  9.76 \\
$\sigma_{ij}$      &  0 &  1 &  1
                   & - $\frac{3}{2}\ln(\mu^2/m_b^2)$
                     + $\frac{3}{2}\ln(a^2 m_b^2)$ - 8.25
                   &  0.56 \\
\end{tabular}
\end{center}
\end{table}

\end{document}